# Band gap and pseudocapacitance of $Gd_2O_3$ doped with $Ni_{0.5}Zn_{0.5}Fe_2O_4$

M. Azeem[1], Q. Abbas[2,3], M. A. Abdelkareem[2], A.G. Olabi[2,4]

[1]Department of Applied Physics and Astronomy, University of Sharjah, 27272, University City, Sharjah, United Arab Emirates.

[2]Sustainable Energy & Power Systems Research Centre, RISE, University of Sharjah, P.O. Box 27272, Sharjah, United Arab Emirates

[3]Institute of Engineering and Energy Technologies (IEET), School of Engineering, Computing & Physical Sciences, University of the West of Scotland, Paisley, PA1 2BE, UK

[4]Mechanical Engineering and Design, Aston University, School of Engineering and Applied Science, Aston Triangle, Birmingham, B4 7ET, UK

**Abstract:**

Herein, we present a detailed study of the structural, optical, and electrochemical responses of $Gd_2O_3$ doped with nickel zinc ferrite nanoparticles. Doping of $Ni_{0.5}Zn_{0.5}Fe_2O_4$ nanoparticles to $Gd_2O_3$ powder was done through thermal decomposition at 1000 °C. The average grain size of the mixture was determined to be approximately 95 nm, and phases of cubic $Gd_2O_3$, GdO, and orthorhombic prisms of $GdFeO_3$ were identified. The focused ion beam energy dispersive X-ray spectrum (FIB-EDX) mapping results clearly show the morphology of the particles with Gd and Fe as the dominant elements. The structural data were compared with the spectroscopic measurements confirming the formation of multiple phases of oxides and ferrites. The measured optical band gap is significantly redshifted to 1.8 eV and is close to that of nitride compounds of gadolinium metal. The measured specific capacitance was almost 7 $Fg^{-1}$ at a current density of 1 $Ag^{-1}$, showing a small drop of 27% when the current density is increased to 10 $Ag^{-1}$. Cyclic voltammetry (CV) plots of the ferrite doped $Gd_2O_3$ electrode at a scan rate of 5 to 100 mV/s indicate the pseudocapacitive nature of the material.

**Introduction**

The $Gd^{3+}$ ion carries a large spin magnetic moment (S=7/2) which makes the gadolinium-based composite materials interesting[1]. They have found their applications in a variety of fields including magnetic coolers[2], high-performance conductors[3], supercapacitors electrodes[4], optoelectronics[5], and magnetic resonance imaging(MRI) [6,7]. A very recent review report has highlighted the spintronic potential of the $Gd^{3+}$ composites in the ferroic-order [ferromagnetic, ferroelectric, and multiferroic] oxides[8]. Gadolinium ferrite based material systems, ($Gd_xFe_{1-x}$/NiCoO) [9] and cobalt gadolinium oxide (Co/$Gd_2O_3$), [10] exhibit a controllable positive exchange bias. Magnetoresistance, coercivity, Curie temperature, magnetic anisotropy, saturated magnetization, and magnetic remanence of several other rare earth oxide based materials systems have also been studied [11,12].

It appears that most of the attention has been focused on the magnetic properties of rare earth oxide-based heterostructures whereas the optical and electrochemical properties have largely been ignored. We know that the fundamental optical absorption edge for $Gd_2O_3$ lies in the range of 4.9 eV to 5.6 eV [13–15] depending on the preparatory methods, crystallite size and the morphology of the prepared materials, from nanoparticles [16] to bulk [13]. These experimentally determined values differ widely with the theoretical calculations [17,18]. There are very limited attempts to show the practicality of these materials systems in devices. A single domain, crystalline $Gd_2O_3$ film grown on Si by molecular beam epitaxy has shown a potential to be used in super capacitors with a leakage current density of about $10^{-3}$ $A/cm^2$[4]. The $Gd_2O_3$ has aslo been successfully employed as an effective component of high-performance composites in number of studies [19,20]. Lanthanide doped metal oxides have shown to improve the specific capacitance. For example, for the case of double perovskites $Gd_2NiMnO_6$, the measured specific capacitance was ~400.46 $Fg^{-1}$ at current density of 1 $Ag^{-1}$ [21] while Gd-doped $HfO_2$ and Aluminium doping in $Sr_3Co_2Fe_{24}O_{41}$ [22,23] are also found to have potential to be used in the supercapacitor application as well. These values of the capacitance are not very far from the transition metal based oxides such as cobalt-doped ZnO quantum dots where the specific capacitance of 697 $Fg^{-1}$ is observed along with power and energy density of around 1026 W/kg and 24Wh/kg, respectively [24]. However, NiCoP/NC electrodes [25] have achieved superb rate-capability, showing capacitance of 1127 Fg$^{-1}$ and 873 Fg$^{-1}$ at current density values of 1 and 16 Ag$^{-1}$, respectively, long-term durability (75.5% retention after 8000 cycles) and ultra-high energy density of around 52.5 Wh kg$^{-1}$.

In this report we make an attempt to unravel the fundamental properties of the ferrite doped Gd$_2$O$_3$ such as the energy band gap and pseudocapacitance relevant to the practical applications. The Gd$_2$O$_3$ was choosen due to its highly stable nature when compared with other rare earth metal due to its half-filled $4f^7$ levels [26]. We shed light on its electronic structure by measuring the optical band gap. The free carrier density is estimated to be of the order $10^{26}$ cm$^{-3}$ which contributes to the conduction of electrons despite its semiconducting nature. Our study shows that ferrite based Gd$_2$O$_3$ heterostructures have great potential for energy-efficient, high density, and non-volatile data storage.

**Materials and methods**

The gadolinium oxide (Gd$_2$O$_3$) and nickel zinc ferrite (Ni$_{0.5}$Zn$_{0.5}$Fe$_2$O$_4$) nanoparticles were purchased from commercial vendors. The average size of Ni$_{0.5}$Zn$_{0.5}$Fe$_2$O$_4$ nanoparticles was around 50 nm. The sample was prepared by mixing 5 g of Gd$_2$O$_3$ with 1 g of Ni$_{0.5}$Zn$_{0.5}$Fe$_2$O$_4$ nanoparticles in a pestle and mortar. The color of the mixture was reddish dark brown which changed to dark gray after thermal decomposition at a temperature of 1000 °C for 12 hours. The phases of the fabricated material and their crystal structures were examined by X-ray diffraction (XRD) patterns using a Bruker D8 Advance X-ray diffractometer. The Cu Kα radiations of energy 8.04 keV, corresponding to an x-ray wavelength of 1.5406 Å were used to irradiate the samples. The patterns were obtained in a 2θ range of 20° to 85° with a step size of 0.02. The phase analysis was performed by using the software QualX. [27]

The Raman spectra were collected by using the Renishaw inVia Raman spectrometer. A site of the specimen was exposed to a laser source with a spot size of 50 mm and a laser power of 1% (14 mW). For the heterogeneous samples reported here, the Raman signals from the different composites overlapped and were indistinguishable. Therefore, up to 45 spectra were collated from different sites of the same sample with the same laser source, and an average was calculated. The average spectra were, then, normalized. The acquisition time for each spectrum was 50 s. The Fourier transform infrared (FTIR) transmittance spectra were obtained by using JASCO FTIR model 6300 type A spectrometer. Transmittance was measured by setting the resolution of the spectrometer to 2 $cm^{-1}$ in the frequency range of 400 $cm^{-1}$ to 2000 $cm^{-1}$.

The morphology of the fabricated material was studied by using a Tescan VEGA XM variable pressure focused ion beam (FIB) scanning electron microscope (SEM). The X-ray spectrometer embedded in the SEM system also provides energy dispersive spectra (EDX) for the localized chemical analysis. The FIB-EDX mapping was done of the cross-sections of selected regions on the samples.

A differential reflectance spectrometer (DRS), model UV-2600i, was used to measure the diffuse reflectance from the sample. Approximately 1 mm thick pellets were prepared to measure the reflectance in the wavelength range of 200 nm to 800 nm (corresponding to the photon energy range of 6.2 eV to 1.5 eV) with a data interval of 0.5 nm. Barium sulfate, with an absorption coefficient almost equal to zero (and reflectance close to 1), was used as a standard to measure relative reflectance.

Electrodes for electrochemical measurements were prepared using carbon in the form of well grinded powder (grind for around 20 mins ). Approximately 80 wt% active carbon material, 10wt% Cabot carbon black as conductivity enhancer and 10 wt% PVDF powder as binder were used for the fabrication of electrodes. The electrode components were mixed with ethanol for 2 hours to form a paste. A syringe is used to drop the paste on a 1 $cm^2$ on the carbon sheets. The carbon sheets are cleaned using $H_2SO_4$ and $HNO_3$. The electrodes were left to dry in oven at 70 ℃ for 24 hours. A standard three-electrode electrochemical test cell was utilized to evaluate the electrochemical characteristics (cyclic voltammetry, galvanostatic charge/discharge, and electrochemical impedance spectroscopy) of synthesized materials using a potentiostat (BioLogic SP-200 Potentiostat, UK). For the three-electrode configuration, prepared materials were utilized as a working electrodes. Platinum plate and Hg/HgO were used as a counter and reference electrode, respectively. A 2 M KOH aqueous solution was used as an electrolyte.

**Results and discussions**

The XRD pattern of the thermally decomposed mixture of $Gd_2O_3$ doped with $Ni_{0.5}Zn_{0.5}Fe_2O_4$ is shown in Figure 1. The average grain size was determined to be approximately 95 nm by using Scherrer equation. The dominant phases are of cubic $Gd_2O_3$ with cubic gadolinium monoxide (GdO) as minority. The strongest peaks at $2\theta \approx 28.56º$, 33.09º, 35.17°, 39.04°, 42.60°, 47.51º, 52.10°, 54.99° and 56.40º are attributed to $Gd_2O_3$ phases with planes oriented in the [222], [312], [411], [332], [413], [440], [611], [514] and [622] directions respectively. Also, there are several weak phases of $Gd_2O_3$ above $2\theta \approx 60º$. The phases of the cubic $Gd_2O_3$ belong to the space group Ia-3 with the lattice parameter $a=10.8120$ Å. The cell volume is approximately 1263.92 Å$^3$.

Phases of gadolinium ferrites ($GdFeO_3$) were identified at $2\theta \approx 20.22º$, 22.95º, 23.17º, 31.87º, 33.53º, 39.8º, 41.2º, 46.95º, and 48.22º, and 59.15º, corresponding to planes oriented along [101], [110], [002], [020], [200], [022], [202], [220], [023] and [204] respectively. Whereas, the peaks at $2\theta \approx 32.91º$, 53.23º, and 59.53º are because of the diffraction from the planes along [112], [131] and [312] directions, respectively, indicative of positive orthorhombic pyramids. The diffraction angles and the corresponding plane orientations suggest that blocky crystals of $GdFeO_3$ exists in the form of orthorhombic prisms of the 1st, 2nd, and 3rd order. Most of the phases of $GdFeO_3$, therefore, belong to the orthorhombic system having a 2-fold axis of rotational symmetry coincident with the *c* axis of the crystal without the presence of a 3-fold axis of rotational symmetry. The cell parameters for the orthorhombic crystal system of $GdFeO_3$ follow the conventional order of *a<b<c* with the values 5.3490 Å, 5.6089 Å and 7.6687 Å, respectively. The cell volume is approximately equal to 230.08 Å$^3$. The axis of elongation is along *c* axis. The space group is P b n m.

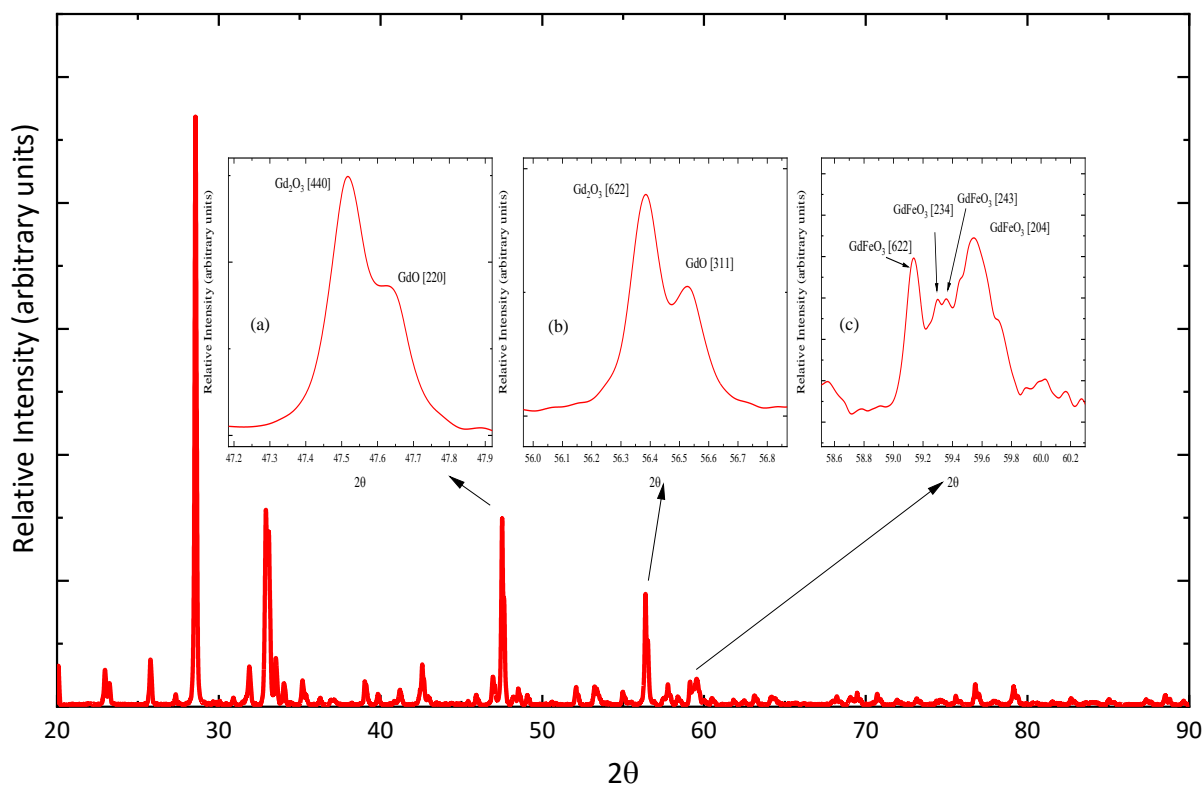

*Figure 1 Experimental XRD pattern. The insets show multiple splitting of the diffracted peaks.*

Further, multiple splitting of the diffraction peaks, inset (a-c) Figure 1, indicate the phase transformations. The split between 47.2º and 47.9º, inset (a) Figure 1, shows two phases of gadolinium oxide identified as, $Gd_2O_3$ and GdO, diffracted from two parallel planes, [440] for $Gd_2O_3$ and [220] for GdO. Both of these phases are also identified between 56.2º and 56.9º, shown in inset (b) of Figure 1, reflecting from parallel planes again, [622] and [311] of $Gd_2O_3$ and GdO respectively. The cubic GdO belongs to the space group F -4 3 m with the cell parameter $a$=5.4000 Å and the cell volume =157.46 Å$^3$. Inset (c) of Figure 1 shows multiple splitting in diffraction peaks of $GdFeO_3$. The twin peaks between 59.2º and 59.4º are contributed by the diffraction from [234] and [243] planes. The more intense peaks are from [622] and [204] directions. It should be particularly noted that the GdO and $GdFeO_3$ have metallic character [18] whereas $Gd_2O_3$ is semiconductor.

Evidence of multiple phases of $Gd_2O_3$ and $GdFeO_3$ phases also comes from Raman spectroscopic measurements as shown in Figure 2. Rare earth sesquioxides, however, are luminescent, and therefore strong non-vibrational luminescent bands quite often hide weak Raman vibrational bands. Therefore, to distinguish the Raman signals from the laser-induced luminescence, the spectra were collated by using two laser sources of different wavelengths, 488 nm and 785 nm. The bands shared by the spectra obtained by both the laser sources are true Raman bands. The fingerprint region for the rare earth alloys lies below 1000 cm$^{-1}$, [28,29] however, we attempted to explore the spectral range of 100 cm$^{-1}$ to 2500 cm$^{-1}$. Moreover, the Raman signals are resolved by using the Lorentz model. Lorentzian peak function has a bell shape and much wider tails than the Gaussian function. According to the dispersion theory, the bands of solids can be modeled with damped harmonic oscillators, which under certain approximations, result in the Lorentz profile.[30] However, there is always some uncertainty in the excitation response and the location of the energy level of the excited state so the duration in which a molecule remains in the excited state influences the width of the Lorentzian line. The Lorentzian peak function is defined by four parameters as follows,

$$f(x) = \frac{2A}{\pi}\left(\frac{w}{4(x-x_c)^2+w^2}\right),$$

where "$x_c$" is the peak position i.e. the center of the peak, "$w$" is full width half maximum (FWHM) and "$A$" is the area under the curve. The height of the Lorentz peak can be determined by,

$$H = \frac{2A}{\pi w}.$$

A summary of the fitting parameters of the Lorentzian peak function along with peak positions is shown in Table 1. The fitting was performed by using a commercial software[31].

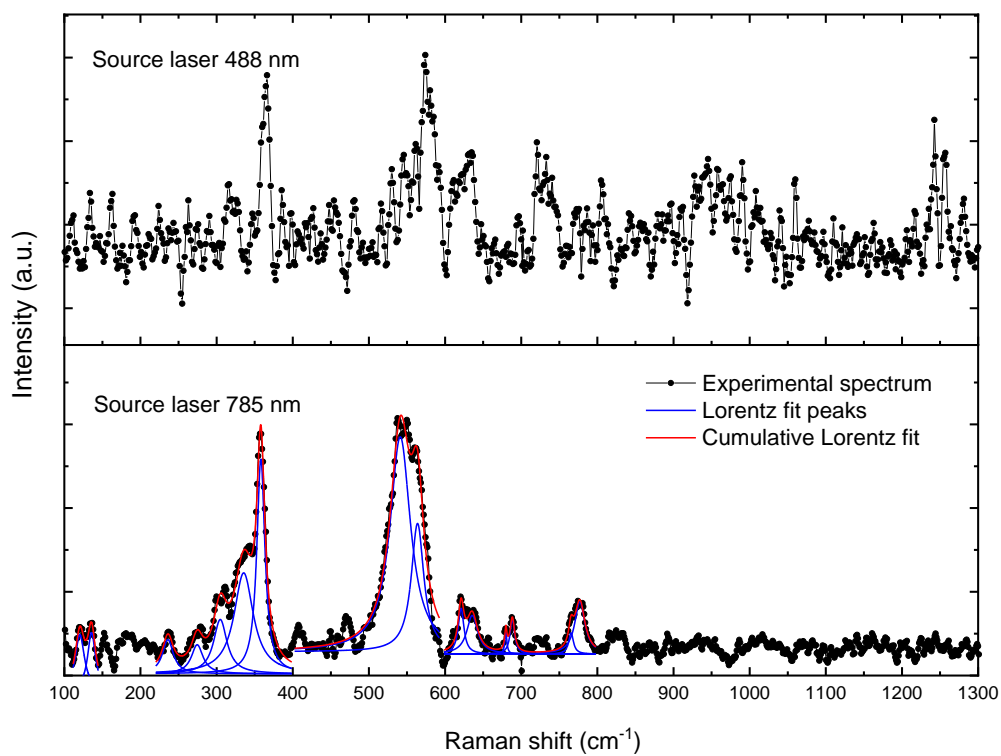

*Figure 2 Raman spectra obtained by laser sources of the wavelengths 488 nm (upper panel) and 785 nm (bottom panel). The common peaks present in both the spectra are true Raman bands.*

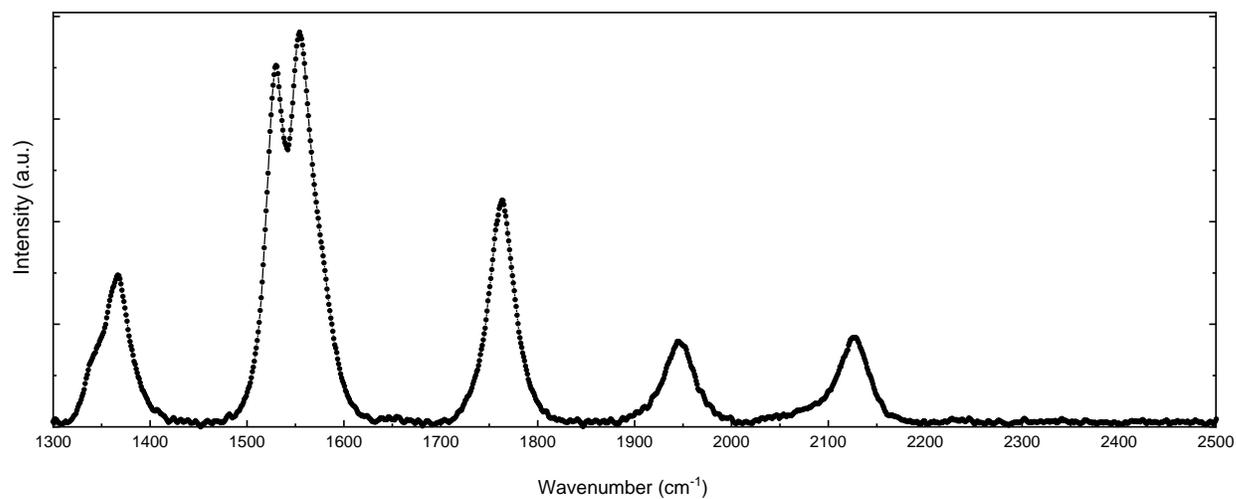

*Figure 3 Flourescence signals induced in Gd$_2$O$_3$ by the laser source of the wavelength 785 nm.*

**Table 1 Fitting parameters for the Lorentzian peak function**

| Plot | Phase | $x_c$ | w | A | $\chi^2$ | $R^2$ |
|---|---|---|---|---|---|---|
| Peak1 | $Gd_2O_3$ | 120.4 ± 0.2 | 14.8 ± 1.2 | 0.6 ± 0.08 | 5.4 x $10^{-6}$ | 0.98 |
| Peak2 | $Gd_2O_3$ | 135.7 ± 0.2 | 12.6 ± 1.04 | 0.5± 0.07 | | |
| Peak3 | $Gd_2O_3$ | 236.0 ± 0.8 | 17.3 ± 2.9 | 0.4 ± 0.07 | | |
| Peak4 | $GdFeO_3$ | 274.5 ± 1.0 | 18.6 ± 4.0 | 0.4 ± 0.08 | | |
| Peak5 | $GdFeO_3$ | 304.8 ± 0.7 | 21.4± 3.1 | 0.8 ± 0.1 | | |
| Peak6 | $Gd_2O_3$ | 335.6 ± 0.6 | 30.5 ± 3.1 | 2.2 ± 0.2 | | |
| Peak7 | $GdFeO_3$ | 358.1 ± 0.1 | 12.8 ± 0.5 | 1.9 ± 0.1 | | |
| Peak9 | $GdFeO_3$ | 541.2 ± 0.6 | 32.7 ± 1.6 | 5.0 ± 0.3 | | |
| Peak10 | $GdFeO_3$ | 563.9 ± 0.6 | 20.0 ± 1.9 | 1.8 ± 0.2 | | |
| Peak11 | $GdFeO_3$ | 621.0 ± 0.4 | 9.0 ± 1.5 | 0.3 ± 0.1 | | |
| Peak12 | $GdFeO_3$ | 635.6 ± 0.9 | 17.4 ± 2.8 | 0.5 ± 0.08 | | |
| Peak13 | $GdFeO_3$ | 679.8 ± 0.5 | 3.4 ± 1.7 | 0.1 ± 0.02 | | |
| Peak14 | $GdFeO_3$ | 688.3 ± 0.5 | 7.3 ± 1.6 | 0.2 ± 0.03 | | |
| Peak15 | $GdFeO_3$ | 765.5 ± 0.9 | 7.4 ± 3.3 | 0.1 ± 0.06 | | |
| Peak16 | $GdFeO_3$ | 777.0 ± 0.8 | 16.3 ± 2.2 | 0.6 ± 0.09 | | |

There are 16 Raman bands, shared by the spectra obtained by 488 nm (upper panel of Figure 2) and 785 nm (lower panel of Figure 2) laser sources. The bands associated with $Gd_2O_3$ are at about 120 $cm^{-1}$, 136 $cm^{-1}$, 236 $cm^{-1}$, 336 $cm^{-1}$, and 564 $cm^{-1}$ whereas vibrational bands located at about 275 $cm^{-1}$, 305 $cm^{-1}$, 358 $cm^{-1}$, 541 $cm^{-1}$, 621 $cm^{-1}$, 636 $cm^{-1}$, 680 $cm^{-1}$, 688 $cm^{-1}$, 766 $cm^{-1}$, and 777 $cm^{-1}$ belong to $GdFeO_3$. Our observations are in close agreement with the previous studies on the similar material systems.[32–35]

Figure 3 shows the spectra in the range of 1300 $cm^{-1}$ to 2500 $cm^{-1}$ obtained by 785 nm laser source where very strong fluorescence peaks can be observed. While the abundance of the fluorescence signals create difficulties in the interpretation of the Raman spectrum, in this case we are able to identify $Gd_2O_3$ by these signals in agreement with the previous report [35].

The SEM micrographs, Figure 4(a,b), show the morphology of the fabricated materials. It appears that the nanoparticles of an average diameter of around 250 nm form micrometer-sized aggregates of $GdFeO_3$ and $Gd_2O_3$. The representative EDX spectra, shown in Figure 5(a,b), were collated from several randomly chosen sites of the samples confirming that the Gd and Fe elements are the highest in concentration followed by Ni and Zn. However, certain sites were completely depleted of the Ni and the Zn atoms as is apparent in Figure 5(b). The FIB-EDX mapping was used to analyze selected particles in the mixture, Figure 5(c). It was observed that the bulk of the particles mainly contains Gd (≈71 wt%), oxygen (≈20 wt%) and Fe (≈7 wt%) as confirmed by the dominant peaks in the EDS spectrum of the particles.

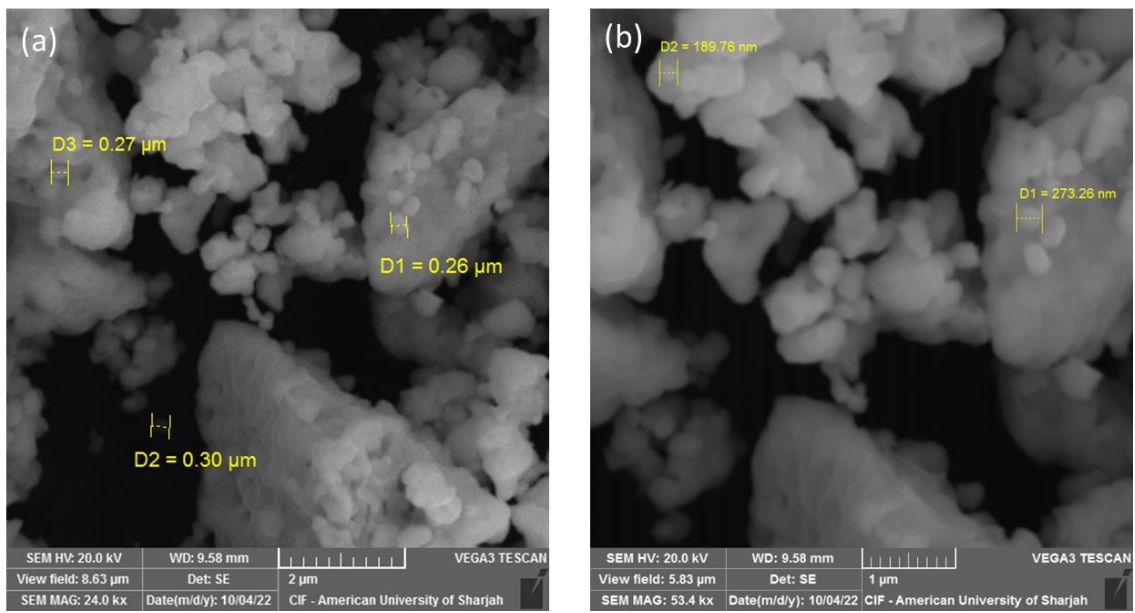

**Figure 4 (a,b) The SEM micrographs from two different randomly selected sites. Micro-sized clusters formed by the aggregation of particles of the average diameter of 250 nm.**

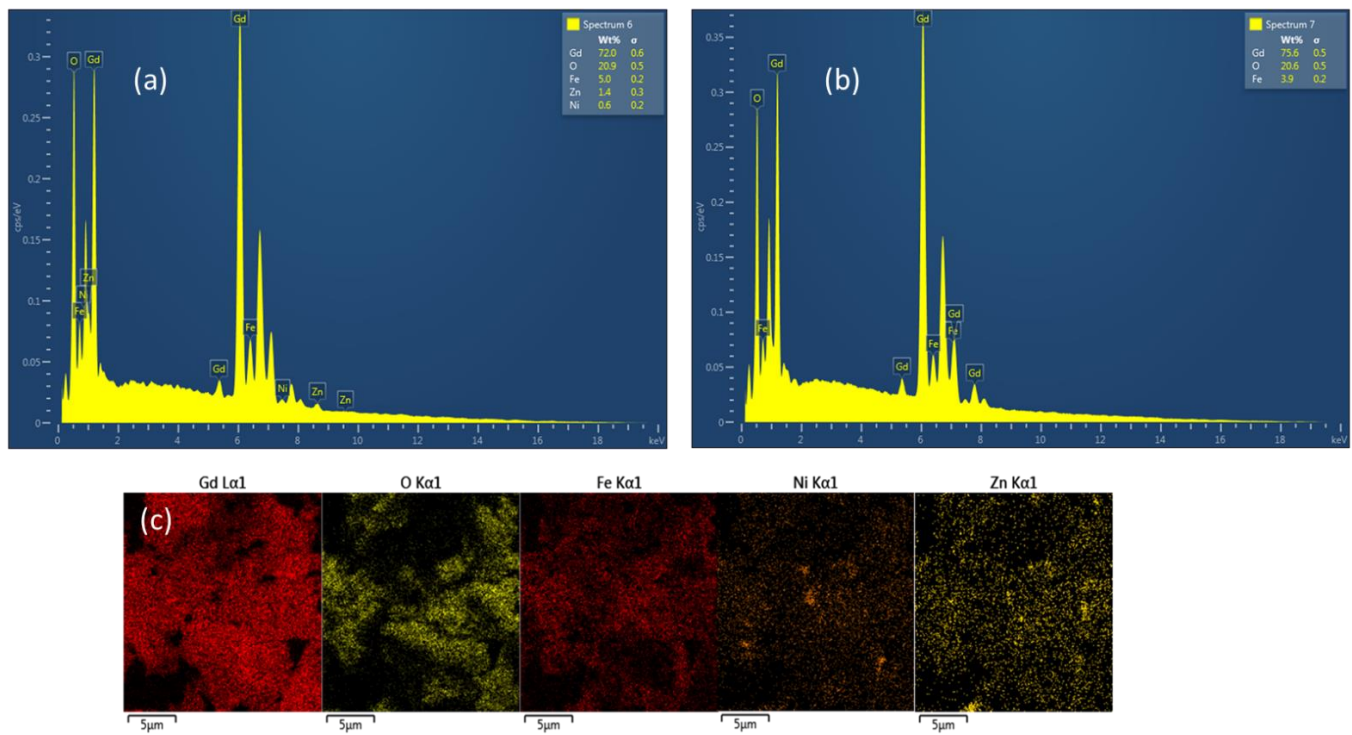

*Figure 5 (a, b) EDX and (c) FIB-EDX of ferrite doped Gd$_2$O$_3$ nanoparticles.*

The FTIR spectrum is shown in Figure 6. In the fingerprint region i.e. below 1600 cm$^{-1}$, the absorption bands at 1508 cm$^{-1}$, 1398 cm$^{-1}$, and 1388 cm$^{-1}$, inset (a) Figure 6, are the characteristic features of gadolinium oxide. Note that the previous studies show only two absorption bands [36] in this spectral range, as opposed to three in this study, indicative of two phases of gadolinium oxide such as $Gd_2O_3$ and $GdO$. Figure 6 inset(b) shows several bands in the spectral range of 600 cm$^{-1}$ to 400 cm$^{-1}$ ascribed to metal-oxides. The band at 547 cm$^{-1}$ is due to the Gd-O vibration of $Gd_2O_3$ whereas the bands from 500 cm$^{-1}$ to 400 cm$^{-1}$ are due to Fe-O and Gd-O vibrations.

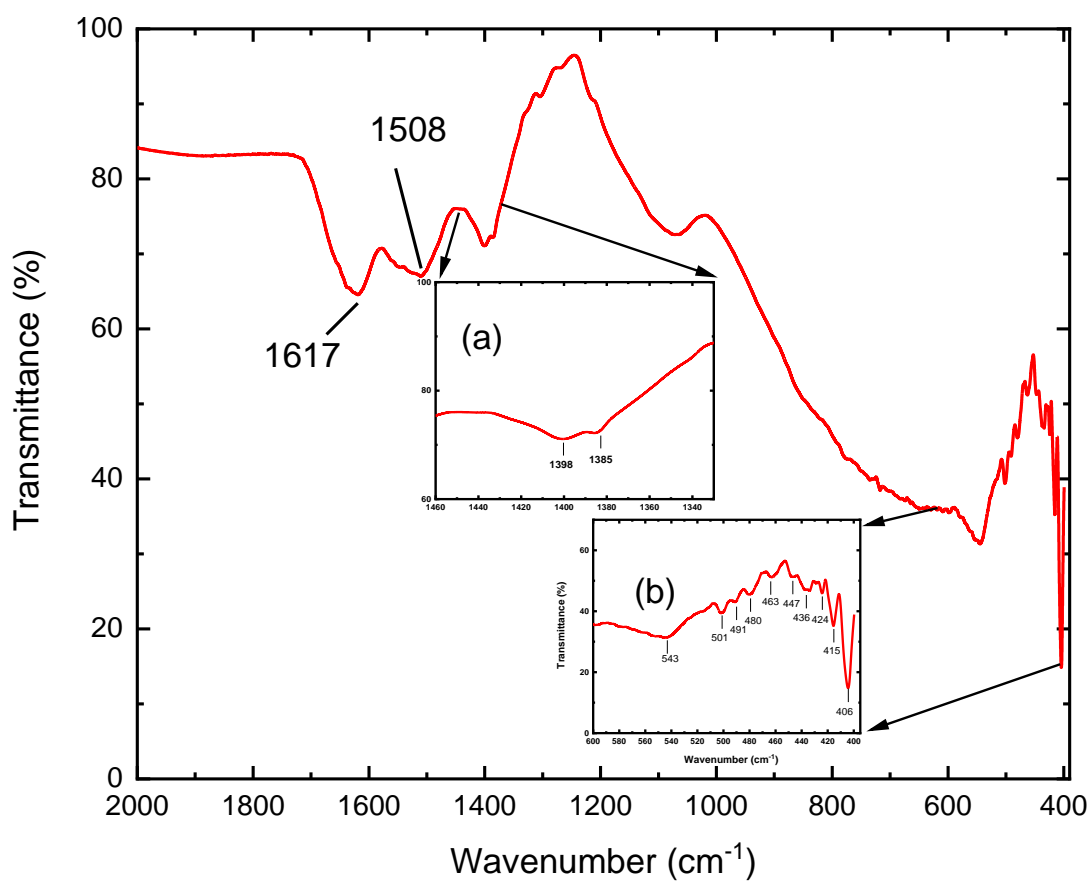

*Figure 6  FTIR spectrum showing absorption bands of the mixed phases. Inset (a) and (b) show zoomed in views in the spectral ranges 1330 cm$^{-1}$ to 1450 cm$^{-1}$ and 400 cm$^{-1}$ to 600 cm$^{-1}$.*

14Diffuse reflectance spectrum and the calculated absorption coefficient is shown Figure 7(a,b) as a function of photon energy. The diffuse reflected light emits after penetrating and repeatedly refracting, transmitting and scattering inside the sample. A monotonic fall in the reflectance spectrum, Figure 7(a), is observed above 1.0 eV to nearly 4.5 eV where it becomes almost flat but then falls again abruptly at about photon energy value of 5.6 eV. This is a clear indication of presence of the optical gap. A rise in the reflectance below 1.0 eV signifies free carriers. The diffuse reflectance is related to the absorption and scattering coefficient by following relationship as put forward by Kubleka-Munk,

$$F(R_\infty) = \frac{K}{S} = \frac{(1 - R_\infty)^2}{2R_\infty},$$

where "$F(R_\infty)$" is referred to as the Kubelka-Munk function, and "$R_\infty$" is the absolute reflectance of the sample, "$K$" is the absorption coefficient, and "$S$" is the scattering coefficient. The scattering coefficient varies depending on the size and density of the particles as well as refractive index of the materials. However, it is not dependent on the light frequency so the scattering coefficient is a constant in the Kubelka-Munk model assuming that refractive index of the absorbing materials is in the range 2.1 – 2.5. [37] The $F(R_\infty)$, therefore, is proportional to the absorption coefficient [38,39]. Considering the refractive index of $Gd_2O_3$ is about 2.1 [40] and the sample pellets of about 1 mm thickness are assumed to be of infinite thickness, [41] the absorption coefficient calculated from Kubelka-Munk model, Figure 5(b), shows an onset of absorption at photon energy ~1.8 eV, which is on average 4.0 eV smaller than the values [13–15,42] reported for $Gd_2O_3$. The rise in the absorption, however, is not monotonic and becomes flat at around 3.8 eV up to 5.6 eV where a second absorption edge lies. A double onset of absorption, at 1.8 eV and 5.6 eV, indicates the formation of a new valence band due to the hybridization of the atomic orbitals of the ferrites and $Gd^{3+}$. It should be noted that the fundamental absorption edge at 1.8 eV is only 0.5 eV greater compared to those of nitrides compounds of rare earth metals [43–45] which are considered to be the leading materials with potential in spintronic applications. A significant absorption in the subgap region is indicative of a high density of electronic charge carriers. The free carrier density can be esitmated by using the Drude model [46] and is found to be of the order $10^{26}$ cm$^{-3}$ which is the characteristic of a heavily doped semiconductor [47]. In a doped state, the charge carriers form donor and acceptor levels, above the valence and below the conduction bands respectively. These delocalized states are responsible for a metallic like electronic transitions in the subgap region.

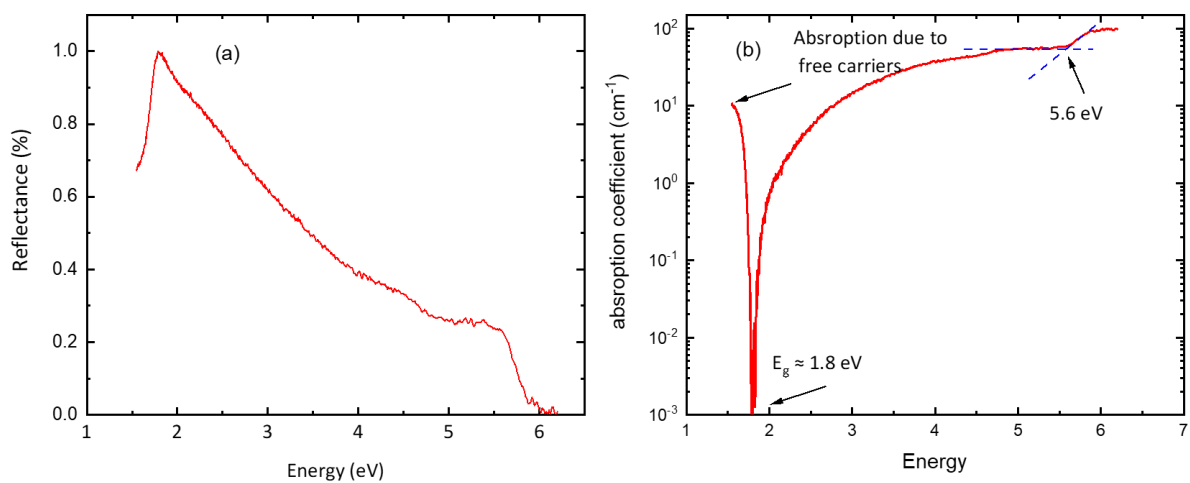

**Figure 7 (a) Diffuse reflectance spectrum shows a fall in the reflectance above 1.8 eV(b) Absorption coefficient calculated by using the Kublek-Munk model from the reflectance spectrum indicates presence of the energy band gap.**

The specific capacitance, in units of F.g$^{-1}$, was calculated using the following equation,

$$\text{Specific capacitance} = \frac{I.\Delta t}{m.\Delta V},$$

where "I" is the discharge current, "Δt" is the discharge time, "ΔV" is the operating potential window, "m" is the mass loading of the active electrode material. A potential window of 0.30 to 0.53 V was used during both the galvanostatic charge-discharge (GCD) and cyclic voltammetry (CV) segments of experiment. Figure 8 shows the GCD measurement profiles at current densities of 1 to 10 Ag$^{-1}$. Specific capacitance was measured from the discharge curve at all current densities and highest capacitance of 7 Fg$^{-1}$ was measured at 1 Ag$^{-1}$. An excellent cyclability is displayed with modest drop of around 27% in capacitive performance when current density was increased ten times from 1 Ag$^{-1}$ to 10 Ag$^{-1}$. The specific capacity values for different current densities are listed in Table 2. The drop in specific capacitance at higher current densities can be credited to the diffusion effect of electrolyte ions on the surface of active material. At higher current densities the inner active sites cannot precede the redox reaction effectively. [48]

Table 2: Specific capacity values of $Gd_2O_3$ at different current densities.

| Current density (A/g) | 1 | 2 | 4 | 6 | 8 | 10 |
|---|---|---|---|---|---|---|
| Specific capacitance (F/g) | 7 | 6 | 5.48 | 5.28 | 5.12 | 5.10 |

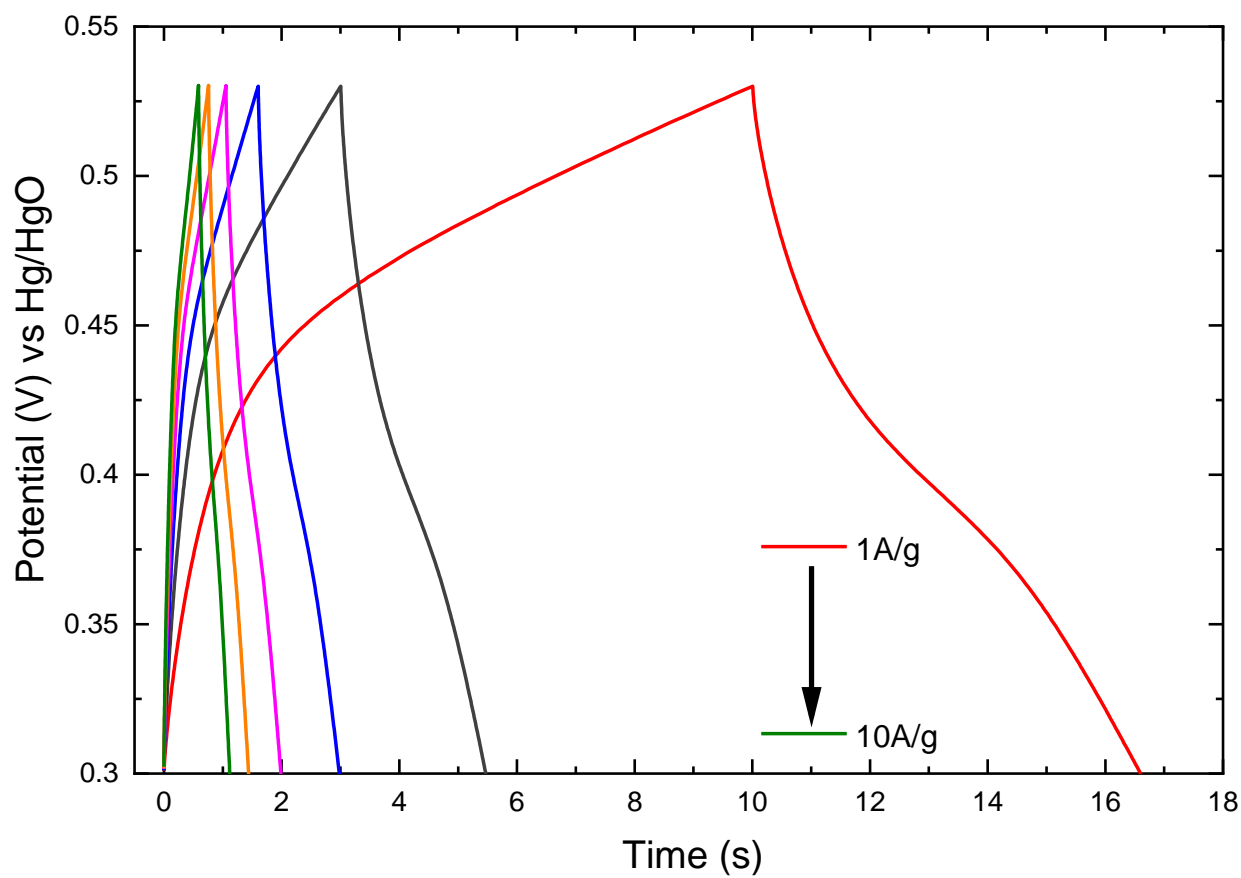

Figure 8 The GCD profile of ferrite doped $Gd_2O_3$ electrodes at current density from 1 to 10 $Ag^{-1}$.

The cyclic voltammetry (CV) plots of ferrite doped $Gd_2O_3$ electrode at the scan rate of 5 mV/s to 100 mV/s is shown in Figure 9. The CV profiles are almost symmetrical in shape within the applied potential window. Well defined oxidation/reduction peaks present in all CV curves represents the pseudocapacitance nature of the tested electrode at all scan rates.

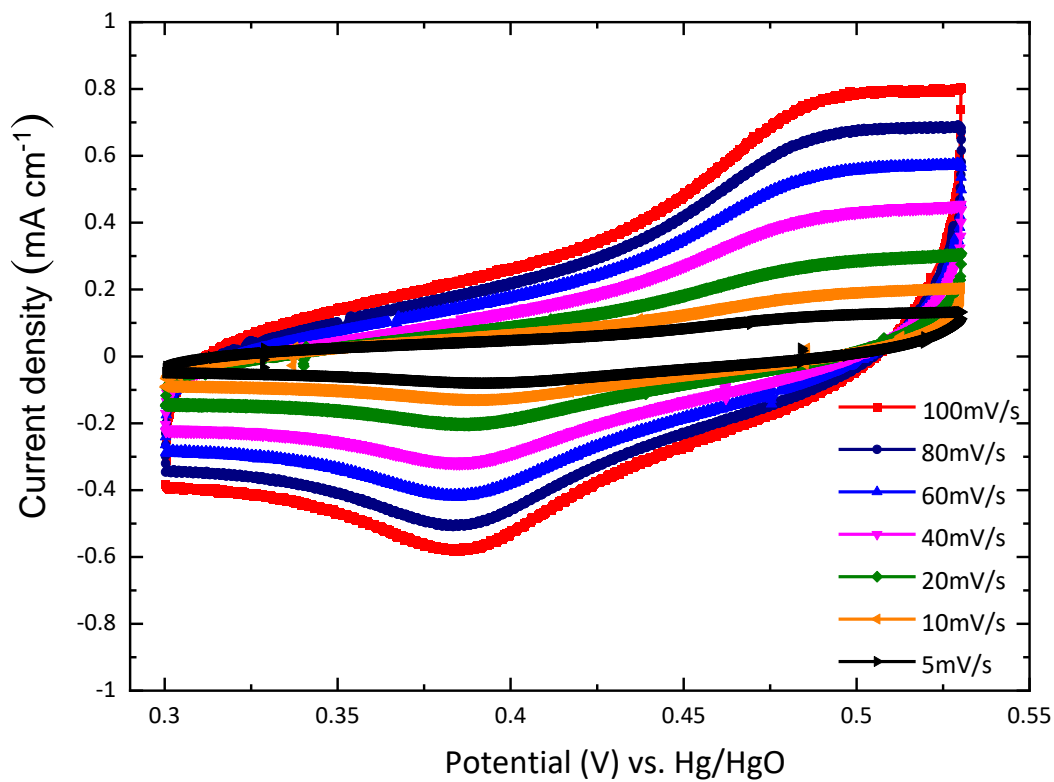

Figure 9 CV profile of $Gd_2O_3$ electrodes at scan rates from 5 to 100 mVs$^{-1}$.

Electrochemical impedance spectroscopy (EIS) is a widely used technique to evaluate the resistive and capacitive behaviour of any active materials for supercapacitor applications. Figure 10(a) shows a typical complex plane impedance plot of ferrite doped $Gd_2O_3$ electrodes where a high frequency intercept of real axis provides an equivalent series resistance (ESR) value, which is the sum of solution and active materials resistances, at around 0.99 Ω as shown in inset of Figure 10(a). The low frequency region of the curves at 45° symbolizes an excellent capacitive behaviour of the electrode material.[49]

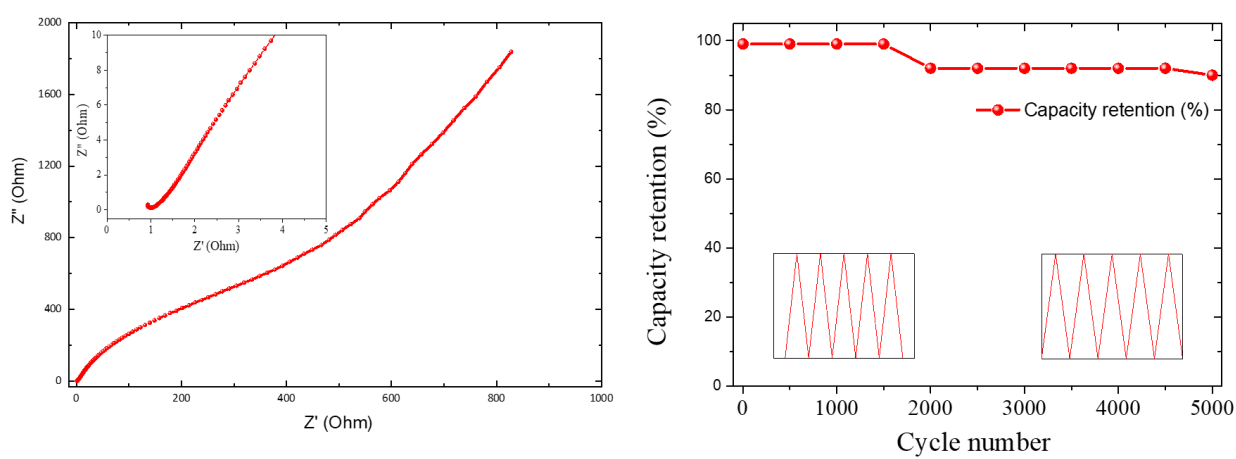

Figure 10 (a) EIS Nyquist plots of $Gd_2O_3$ electrodes in frequency range of 10 mHz to 100 kHz with wave amplitude of 5 mV. (b) Cyclic stability of Gadolinium oxide (inset: initial and final five charge–discharge cycle).

Material stability is another vital aspect of any active material for its potential of real word applications. Gadolinium oxide based active material showed excellent long-term durability with the capacity retention of around 90% after 5,000 cycles as shown in Figure 10(b) where inset shows first and last five charge discharge cycles. This exceptional retention performance can be due to the robustness nature of this material and good conductivity (low resistance) as shown by EIS plot, Figure 10 (a).

We finally show the energy density as a function of power density as shown in the Figure 11. The specific energy density of 0.98 W h /kg at the specific power density of 504 W/kg and an energy density of 0.70 W h /kg was retained at higher power density 5100 W/kg. It can be witnessed that there was slight drop in energy density when power density was increased by over ten folds.

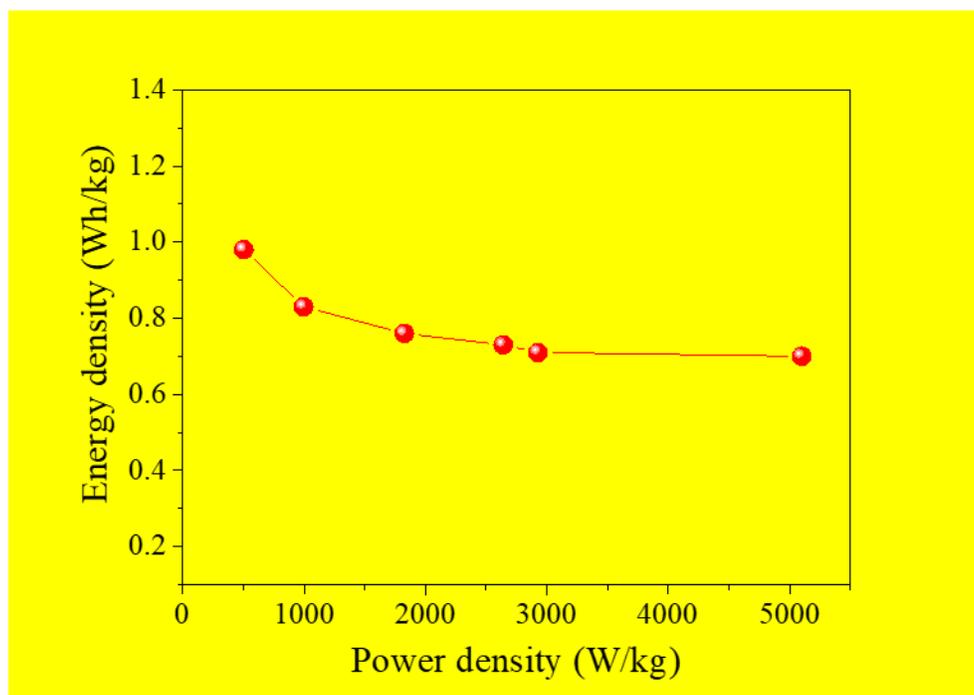

Figure 11: Ragone plot illustrating energy density as a function of power density of Gd2O3 doped with $Ni_{0.5}Zn_{0.5}Fe_2O_4$ based super capacitor.

20**Conclusion**

We have reported a clear experimental signature of the direct optical energy gap in the ferrite doped $Gd_2O_3$. The nanostructures of the ferrite doped $Gd_2O_3$ with grain size of approximately 95 nm were prepared by thermal decomposition. The fundamental energy band gap appeared to be red shifted at 1.8 eV, as determined by absorption coefficient using the Kubleka-Munk model. The highest specific capacitance was measured to be of 7 Fg$^{-1}$ at current density of at 1 Ag$^{-1}$ which showed a drop of only 27% as the current density was increased. Although, the drop in the capacitance is small, the specific capacitance itself, however, is of modest value. A comparative CV curves obtained for different scan rates illustrates the pesudocapacitve nature of the ferrite doped $Gd_2O_3$. The techniques employed in this study show that $Gd_2O_3$ doped with 20 wt% concentration of $Ni_{0.5}Zn_{0.5}Fe_2O_4$ exhibits an improved electrochemical behaviors which may be used for high-performance pseudocapacitor applications. Ferrite doped $Gd_2O_3$ heterostructures synthesized with simple approach of thermal decomposition display superior electrode behaviours and could be used to fabricate more electroactive electrode materials.

**Acknowledgement** The work is supported by the University of Sharjah competitive research grant number 21021430112. Authors are thankful to Anakha Udayan for data acquisition.**Ethical Compliance** All procedures performed in studies involving human participants were in accordance with the ethical standards of the institutional and/or national research committee and with the 1964 Helsinki Declaration and its later amendments or comparable ethical standards.

**Conflict of Interest declaration** The authors declare that they have NO affiliations with or involvement in any organization or entity with any financial interest in the subject matter or materials discussed in this manuscript.

**References**

[1] Lei X gong, Jockusch S, Turro N J, Tomalia D A and Ottaviani M F 2008 EPR characterization of gadolinium(III)-containing-PAMAM-dendrimers in the absence and in the presence of paramagnetic probes *J. Colloid Interface Sci.* **322** 457–64

[2] Sibille R, Didelot E, Mazet T, Malaman B and François M 2014 Magnetocaloric effect in gadolinium-oxalate framework Gd2(C2O4)3(H2O)6·(0.6H2O) *APL Mater.* **2**

[3] MacManus-Driscoll J L, Bianchetti M, Kursumovic A, Kim G, Jo W, Wang H, Lee J H, Hong G W and Moon S H 2014 Strong pinning in very fast grown reactive co-evaporated GdBa 2Cu3O7 coated conductors *APL Mater.* **2**

[4] Wilk G D, Wallace R M and Anthony J M 2001 High-κ gate dielectrics: Current status and materials properties considerations *J. Appl. Phys.* **89** 5243–75

[5] Hazarika S and Mohanta D 2013 Production and optoelectronic response of Tb3+ activated gadolinium oxide nanocrystalline phosphors *EPJ Appl. Phys.* **62** 3–8

[6] Alizadeh M J, Kariminezhad H, Monfared A S, Mostafazadeh A, Amani H, Niksirat F and Pourbagher R 2019 An experimental study about the application of Gadolinium oxide nanoparticles in magnetic theranostics *Mater. Res. Express* **6**

[7] Sakai N, Zhu L, Kurokawa A, Takeuchi H, Yano S, Yanoh T, Wada N, Taira S, Hosokai Y, Usui A, Machida Y, Saito H and Ichiyanagi Y 2012 Synthesis of Gd2O3 nanoparticles for MRI contrast agents *J. Phys. Conf. Ser.* **352**

[8] Gu Y, Song C, Wang Q, Hu W, Liu W, Pan F and Zhang Z 2021 Emerging opportunities for

voltage-driven magneto-ionic control in ferroic heterostructures *APL Mater.* **9**

[9]     Gilbert D A, Olamit J, Dumas R K, Kirby B J, Grutter A J, Maranville B B, Arenholz E, Borchers J A and Liu K 2016 Controllable positive exchange bias via redox-driven oxygen migration *Nat. Commun.* **7**

[10]    Bi C, Liu Y, Newhouse-Illige T, Xu M, Rosales M, Freeland J W, Mryasov O, Zhang S, te Velthuis S G E and Wang W G 2014 Reversible Control of Co Magnetism by Voltage-Induced Oxidation *Phys. Rev. Lett.* **113** 267202

[11]    Gilbert D A, Grutter A J, Murray P D, Chopdekar R V., Kane A M, Ionin A L, Lee M S, Spurgeon S R, Kirby B J, Maranville B B, N'Diaye A T, Mehta A, Arenholz E, Liu K, Takamura Y and Borchers J A 2018 Ionic tuning of cobaltites at the nanoscale *Phys. Rev. Mater.* **2** 104402

[12]    Yan Y N, Zhou X J, Li F, Cui B, Wang Y Y, Wang G Y, Pan F and Song C 2015 Electrical control of Co/Ni magnetism adjacent to gate oxides with low oxygen ion mobility *Appl. Phys. Lett.* **107** 122407

[13]    Losovyj Y B, Wooten D, Santana J C, An J M, Belashchenko K D, Lozova N, Petrosky J, Sokolov A, Tang J, Wang W, Arulsamy N and Dowben P A 2009 Comparison of n-type Gd2O3 and Gd-doped HfO 2 *J. Phys. Condens. Matter* **21**

[14]    Manigandan R, Giribabu K, Suresh R, Vijayalakshmi L, Stephen A and Narayanan V 2013 Structural, optical and magnetic properties of gadolinium sesquioxide nanobars synthesized via thermal decomposition of gadolinium oxalate *Mater. Res. Bull.* **48** 4210–5

[15]   Zatsepin A F, Kuznetsova Y A, Rychkov V N and Sokolov V I 2017 Characteristic features of optical absorption for Gd2O3 and NiO nanoparticles *J. Nanoparticle Res.* **19**

[16]   Zatsepin A F and Review A 2017 Optical properties and energy parameters of Gd 2 O 3 and Gd 2 O 3 : Er nanoparticles Optical properties and energy parameters of Gd 2 O 3 and Gd 2 O 3 : Er nanoparticles 2–6

[17]   Zhang F X, Lang M, Wang J W, Becker U and Ewing R C 2008 Structural phase transitions of cubic Gd2 O3 at high pressures *Phys. Rev. B - Condens. Matter Mater. Phys.* **78**

[18]   Jain A, Ong S P, Hautier G, Chen W, Richards W D, Dacek S, Cholia S, Gunter D, Skinner D, Ceder G and Persson K A 2013 Commentary: The Materials Project: A materials genome approach to accelerating materials innovation *APL Mater.* **1** 011002

[19]   Boopathi G, Karthikeyan G G, Jaimohan S M, Pandurangan A and De Barros A L F 2018 Dopant Effects of Gd3+ on the Electrochemical Pseudocapacitive Characteristics of Electroactive Mesoporous NiO Electrodes for Supercapacitors *J. Phys. Chem. C* **122** 9257–74

[20]   Poudel M B and Kim H J 2021 Synthesis of high-performance nickel hydroxide nanosheets/gadolinium doped-α-MnO2 composite nanorods as cathode and Fe3O4/GO nanospheres as anode for an all-solid-state asymmetric supercapacitor *J. Energy Chem.* **64** 475–84

[21]   Kumar A, Kumar A and Kumar A 2020 Energy storage properties of double perovskites Gd 2 NiMnO 6 for electrochemical supercapacitor application **105**

[22]   Skopin E V., Guillaume N, Alrifai L, Gonon P and Bsiesy A 2022 Sub-10-nm ferroelectric Gd-doped HfO2layers *Appl. Phys. Lett.* **120**

[23]   Mishra S, Das A and Akhtar A J 2021 Room temperature tuning of non volatile magnetoelectric memory in Al doped Sr3Co2Fe24O41 *Ceram. Int.* **47** 29261–6

[24]   Dutta A, Chatterjee K, Mishra S, Saha S K and Akhtar A J 2022 An insight into the electrochemical performance of cobalt-doped ZnO quantum dot for supercapacitor applications *J. Mater. Res.*

[25]   Zhao T, Liu C, Yi F, Liu X, Gao A, Shu D and Ling J 2021 Promoting high-energy supercapacitor performance over NiCoP/N-doped carbon hybrid hollow nanocages via rational architectural and electronic modulation *Appl. Surf. Sci.* **569** 151098

[26]   Xu A W, Gao Y and Liu H Q 2002 The Preparation, Characterization, and their Photocatalytic Activities of Rare-Earth-Doped TiO2 Nanoparticles *J. Catal.* **207** 151–7

[27]   Altomare A, Corriero N, Cuocci C, Falcicchio A, Moliterni A, Rizzi R and IUCr 2015 *QUALX2.0* : a qualitative phase analysis software using the freely available database POW_COD *J. Appl. Crystallogr.* **48** 598–603

[28]   Tang P, Kuang D, Yang S and Zhang Y 2016 Structural, morphological and multiferroic properties of the hydrothermally grown gadolinium (Gd) and manganese (Mn) doped sub-micron bismuth ferrites *J. Alloys Compd.* **656** 912–9

[29]   Puli V S, Adireddy S and Ramana C V. 2015 Chemical bonding and magnetic properties of gadolinium (Gd) substituted cobalt ferrite *J. Alloys Compd.* **644** 470–5

[30]    Yuan X and Mayanovic R A 2017 An Empirical Study on Raman Peak Fitting and Its Application to Raman Quantitative Research *Appl. Spectrosc.* **71** 2325–38

[31]    Originlab Corporation 2019 Origin (Pro)

[32]    Dilawar N, Varandani D, Mehrotra S, Poswal H K, Sharma S M and Bandyopadhyay A K 2008 Anomalous high pressure behaviour in nanosized rare earth sesquioxides *Nanotechnology* **19**

[33]    Sarkar B J, Deb A K and Chakrabarti P K 2016 XRD, HRTEM, Raman and magnetic studies on chemically prepared nanocrystalline Fe-doped gadolinium oxide (Gd1.90Fe0.10O3-δ) annealed in vacuum *RSC Adv.* **6** 6395–404

[34]    Sena N C, Castro T J, Garg V K, Oliveira A C, Morais P C and da Silva S W 2017 Gadolinium ferrite nanoparticles: Synthesis and morphological, structural and magnetic properties *Ceram. Int.* **43** 4042–7

[35]    Yu J, Cui L, He H, Yan S, Hu Y and Wu H 2014 Raman spectra of RE2O3 (RE=Eu, Gd, Dy, Ho, Er, Tm, Yb, Lu, Sc and Y): Laser-excited luminescence and trace impurity analysis *J. Rare Earths* **32** 1–4

[36]    Chen F, Zhang X H, Hu X D, Zhang W, Zeng R, Liu P D and Zhang H Q 2016 Synthesis and characteristics of nanorods of gadolinium hydroxide and gadolinium oxide *J. Alloys Compd.* **664** 311–6

[37]    Simmons E L 1975 Diffuse reflectance spectroscopy: a comparison of the theories *Appl. Opt.* **14** 1380

[38]   Kumar P and Vedeshwar A G 2018 Mapping the conduction band edge density of states of γ-In2Se3 by diffuse reflectance spectra *J. Appl. Phys.* **123**

[39]   Yang L and Miklavcic S J 2005 Revised Kubelka–Munk theory III A general theory of light propagation in scattering and absorptive media *J. Opt. Soc. Am. A* **22** 1866

[40]   Medenbach O, Dettmar D, Shannon R D, Fischer R X and Yen W M 2001 Refractive index and optical dispersion of rare earth oxides using a small-prism technique *J. Opt. A Pure Appl. Opt.* **3** 174–7

[41]   Danckwerts P V 1962 Angewandte chemie *Chem. Eng. Sci.* **17** 955

[42]   Jamnezhad H and Jafari M 2016 Structure of Gd2O3 nanoparticles at high temperature *J. Magn. Magn. Mater.* **408** 164–7

[43]   Azeem M 2018 On the optical energy gap of SmN *Chinese J. Phys.* **56**

[44]   Azeem M, Ruck B J, Do Le B, Warring H, Trodahl H J, Strickland N M, Koo A, Goian V and Kamba S 2013 Optical response of DyN *J. Appl. Phys.* **113**

[45]   Trodahl H J, Preston A R H, Zhong J, Ruck B J, Strickland N M, Mitra C and Lambrecht W R L 2007 Ferromagnetic redshift of the optical gap in GdN *Phys. Rev. B - Condens. Matter Mater. Phys.* **76**

[46]   Azeem M 2019 Quantitative measure of nitrogen vacancy related effects in SmN and EuN *Adv. Nat. Sci. Nanosci. Nanotechnol.* **10** 015003

[47]   Azeem M 2017 Spin polarized conduction and valence band states in GdN *MRS Adv.* **2** 153–8

[48]   Pandit B, Bommineedi L K and Sankapal B R 2019 Electrochemical engineering approach of high performance solid-state flexible supercapacitor device based on chemically synthesized VS2 nanoregime structure *J. Energy Chem.* **31** 79–88

[49]   Ehsani A 2016 Inhibitory effect of new oxazol derivative on corrosion of stainless steel in acidic medium: An electrochemical investigation *Indian J. Chem. Technol.* **23** 289–95

[50]   Dhanalakshmi S, Mathi Vathani A, Muthuraj V, Prithivikumaran N and Karuthapandian S 2020 Mesoporous Gd2O3/NiS2 microspheres: a novel electrode for energy storage applications *J. Mater. Sci. Mater. Electron.* **31** 3119–29

[51]   Shiri H M and Ehsani A 2016 Pulse electrosynthesis of novel wormlike gadolinium oxide nanostructure and its nanocomposite with conjugated electroactive polymer as a hybrid and high efficient electrode material for energy storage device *J. Colloid Interface Sci.* **484** 70–6